# Some conclusive considerations on the comparison of the ICARUS $\nu_\mu \to \nu_e$ oscillation search with the MiniBooNE low-energy event excess

M. Antonello[1], B. Baibussinov[2], P. Benetti[3], F. Boffelli[3], A. Bubak[11], E. Calligarich[3], S. Centro[2], A. Cesana[4], K. Cieslik[5], D. B. Cline[6], A.G. Cocco[7], A. Dabrowska[5], A. Dermenev[8], A. Falcone[3], C. Farnese[2], A. Fava[2], A. Ferrari[9], D. Gibin[2], S. Gninenko[8], A. Guglielmi[2], M. Haranczyk[5], J. Holeczek[11], M. Kirsanov[8], J. Kisiel[11], I. Kochanek[11], J. Lagoda[10], S. Mania[11], A. Menegolli[3], G. Meng[2], C. Montanari[3], S. Otwinowski[6], P. Picchi[12], F. Pietropaolo[2], P. Plonski[14], A. Rappoldi[3], G.L. Raselli[3], M. Rossella[3], C. Rubbia[1,9,13], P. Sala[4], A. Scaramelli[4], F. Sergiampietri[15], D. Stefan[4], R. Sulej[10], M. Szarska[5], M. Terrani[4], M. Torti[3], F. Varanini[2], S. Ventura[2], C. Vignoli[1], H. Wang[6], X. Yang[6], A. Zalewska[5], A. Zani[3], K. Zaremba[14]

(ICARUS Collaboration)

[1] *INFN - Laboratori Nazionali del Gran Sasso, Assergi, Italy*
[2] *Dipartimento di Fisica e Astronomia Università di Padova and INFN, Padova, Italy*
[3] *Dipartimento di Fisica Università di Pavia and INFN, Pavia, Italy*
[4] *INFN, Milano, Italy*
[5] *H. Niewodniczanski Institute of Nuclear Physics, Polish Academy of Science, Krakow, Poland*
[6] *Department of Physics and Astronomy, UCLA, Los Angeles, USA*
[7] *Dipartimento di Scienze Fisiche Università Federico II di Napoli and INFN, Napoli, Italy*
[8] *INR RAS, Moscow, Russia*
[9] *CERN, Geneva, Switzerland*
[10] *National Centre for Nuclear Research, Otwock/Swierk, Poland*
[11] *Institute of Physics, University of Silesia, Katowice, Poland*
[12] *INFN Laboratori Nazionali di Frascati, Frascati, Italy*
[13] *GSSI, L'Aquila, Italy*
[14] *Institute of Radioelectronics, Warsaw University of Technology, Warsaw, Poland*
[15] *INFN, Pisa, Italy*
   *E-mail*: daniele.gibin@pd.infn.it

ABSTRACT: A sensitive search for anomalous LSND-like $\nu_\mu$ to $\nu_e$ oscillations has been performed by the ICARUS Collaboration exposing the T600 LAr-TPC to the CERN to Gran Sasso (CNGS) neutrino beam. The result is compatible with the absence of additional anomalous contributions giving a limit to oscillation probability of $3.4 \times 10^{-3}$ and $7.6 \times 10^{-3}$ at 90% and 99% confidence levels respectively showing a tension between these new limits and the low-energy event excess ($200 < E_\nu^{QE} < 475$ MeV) reported by MiniBooNE Collaboration. A more detailed comparison of the ICARUS data with the MiniBooNE low-energy excess has been performed, including the energy resolution as obtained from the official MiniBooNE data release. As a result the previously reported tension is confirmed at 90% C.L., suggesting an unexplained nature or an otherwise instrumental effect for the MiniBooNE low energy event excess.

# Contents

## 1. Introduction

The ICARUS Collaboration performed a sensitive search for anomalous LSND like $\nu_\mu$ to $\nu_e$ oscillations with the T600 LAr-TPC exposed to CERN to Gran Sasso (CNGS) neutrino beam [1]. The analysis was based on a collected sample of 1995 neutrino interactions, which corresponds to a $6\times10^{19}$ proton on target statistics. Four clear $\nu_e$ events have been visually identified over the full sample, compared with an expectation of 6.4 ± 0.9 events from conventional sources. The result is compatible with the absence of additional anomalous contributions giving a limit to oscillation probability of $3.4\times10^{-3}$ and $7.6\times10^{-3}$ at 90% and 99% confidence levels respectively.

The MiniBooNE Collaboration observed a new effect in the energy region $200 < E_\nu^{QE} < 475$ MeV — below the sensitive E/L region of LSND — and a significant additional anomaly has been reported [2] both for neutrino and antineutrino data. Tension between the limit $\sin^2(2\theta_{new}) < 6.8\times10^{-3}$ at 90% CL and $< 1.52\times10^{-2}$ at 99% CL of the ICARUS experiment and the neutrino lowest energy points of MiniBooNE ($200 < E_\nu^{QE} < 475$ MeV), was observed suggesting an instrumental or otherwise unexplained nature of the low energy signal reported by Ref. [2]. A similar search performed at the same CNGS beam by the OPERA experiment confirmed ICARUS result and the absence of anomalous oscillations with an independent limit $\sin^2(2\theta_{new}) < 7.2\times10^{-3}$ [3].

The ICARUS analysis method [1] adopted for the comparison with the MiniBooNE results was strongly criticized by MiniBooNE Collaboration in a recent paper [4]. An updated and revised comparison between the ICARUS limit and the MiniBooNE neutrino data, based on the MiniBooNE official data release [5], is presented, with a particular emphasis on the low energy event excess as recorded by MiniBooNE.

## 2. Addressing the MiniBooNE event excess.

As clearly stated in Ref. [1], the ICARUS analysis started from figure 2 bottom of Ref. [2] where the $\nu_e$ CCQE event excess observed by MiniBooNE is shown together with its estimated error in bins of reconstructed neutrino energy $E_\nu^{QE}$. In the same figure the expected event excess $\nu_{e\_osc}$ ($\sin^2(2\theta)$, $\Delta m^2$) for a two-neutrino oscillation interpretation for a set of oscillation parameters ($\sin^2(2\theta)$, $\Delta m^2$) is also shown. The excess of $\nu_e$ CCQE events in each bin of $E_\nu^{QE}$ in the case of full oscillation probability was obtained in [1] by dividing the event excess $\Delta\nu_e$ by the oscillation probability for $\sin^2(2\theta) = 0.2$, $\Delta m^2 = 0.1\text{eV}^2$ computed at the center of each energy bin. Finally the oscillation probability and its corresponding error shown in [1] (and reproduced in Figure 1) were obtained as the ratio between the observed excess $\Delta\nu_e$ of $\nu_e$ CCQE events and the corresponding expected excess of $\nu_e$ CCQE in case of full oscillation $\nu_{e\_osc}(P_{osc}=1)$ as a function of $E_\nu/L$, where $E_\nu$ is the neutrino energy and L is the neutrino propagation distance.

The ICARUS analysis was implicitly assuming that: a) no systematic difference is present between the true neutrino energy ($E_{TRUE}$) and $E_\nu^{QE}$ and b) the bin size is scaled to the energy resolution so that no severe smearing between different energy bins is present. Under such hypotheses the ICARUS treatment would provide correct results.

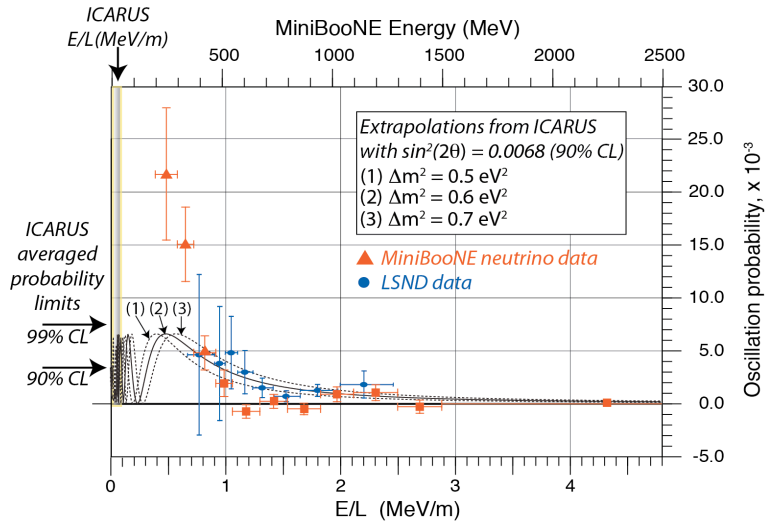

**Figure 1. Oscillation probability limits coming from ICARUS [1] compared with corresponding data from neutrinos in MiniBooNE [2] as a function of $E_\nu/L$ (this picture is a reproduced from [1]). Figure 2 in Ref. [2] has been used in order to convert the observed number of excess events/MeV to their corresponding oscillation probabilities. In order to perform the conversion, the values $\sin^2(2\theta) = 0.2$ and $\Delta m^2_{41} = 0.1$ eV$^2$ have been used. The resulting oscillation probability distribution for neutrino with $E_\nu > 475$ MeV (square red points) appears incompatible with the antineutrino LNSD effect. In the $200 < E_\nu^{QE} < 475$ MeV region (triangular red points) — below the sensitive $E/L$ region of LSND — the new MiniBooNE effect is widely incompatible with the averaged upper probability limits to anomalies from ICARUS [1] and from OPERA [3] in their L/E regions. An extrapolation from ICARUS (black curves marked as 1, 2 and 3) to larger values of E/L for two-neutrino oscillation parameters simultaneously compatible with LSND, MiniBooNE and Karmen is also shown as guidance.**

However some objections to this procedure have been raised by a recent paper by the MiniBooNE Collaboration [4]. The key point is the large difference in MiniBooNE data between the true neutrino energy ($E_{TRUE}$) and the corresponding energy ($E_\nu^{QE}$) reconstructed with "the measured energy and angle of the outgoing muon or electron assuming charged-current quasi-elastic (CCQE) kinematics for the event" [2]. It appears that the reconstructed energy is affected by a huge non-Gaussian smearing compared with the true neutrino energy, as clearly stated in [4] (see Figure 2), in contrast with the much better 11% resolution on $\nu_e$ event energy quoted in a previous paper [6]. This difference between $E_{TRUE}$ and $E_\nu^{QE}$, for which

– 2 –

MiniBooNE gave a quite elliptical explanation[1], is the major cause of the problem in using the L/E (or E/L) to compare data with expectations, since:

- the average value of the reconstructed $E_\nu^{QE}$ does not correspond at all to the average true neutrino energy $E_{TRUE}$ in each $E_\nu^{QE}$ bin as shown in table III and IV of Ref. [4] where $<E_\nu^{QE}>$ and $<E_{TRUE}>$ (or $<L/E_\nu^{QE}>$ and $<L/E_{TRUE}>$) are compared bin by bin;
- the smearing around the average value is very large, so that the neutrinos contributing to each $E_\nu^{QE}$ bin span a very large $E_{TRUE}$ region.

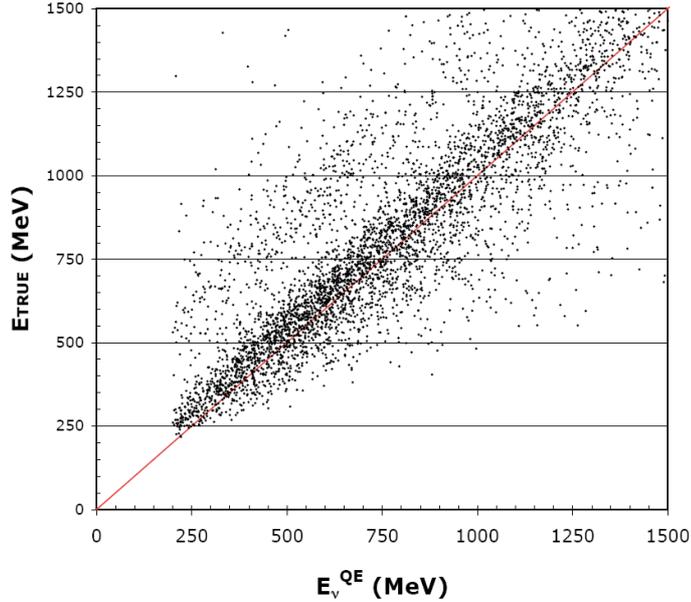

**Figure 2. Energy distribution for neutrino-mode: $E_{TRUE}$ versus $E_\nu^{QE}$ as obtained from Ref. [4] assuming that all muon neutrino are oscillated to $\nu_e$.**

The difference between $E_{TRUE}$ and $E_\nu^{QE}$ can be better evaluated from the MiniBooNE data release [5], studying the spectrum of the true neutrino energy contributing to the individual $E_\nu^{QE}$ bins. Figure 3 shows the distributions of $E_{TRUE}$ for the first four $E_\nu^{QE}$ bins determined assuming that all muon neutrino are oscillated to $\nu_e$ ("fully oscillated sample"). It is evident that $E_\nu^{QE}$ is systematically different from $E_{TRUE}$. Moreover each reconstructed energy bin only marginally overlaps with the corresponding $E_{TRUE}$ bin and gets contributions from many different neutrino energy bins, resulting in a large smearing which can be quantified by the rms width of each experimental point in $L/E_{TRUE}$, (see Table I)[2]. The corresponding error bars on $L/E_{TRUE}$ are shown in Figure 4 as obtained starting from paper [4]. The individual measurements are completely interlaced with one another, as already shown in Figure 3. The chosen energy binning is much finer than the MiniBooNE experimental energy resolution as present in the official data release [5].

---

[1] Both the bias and the long tail are attributed to the same reason: *"there can be a sizable shift in the mean due to the contamination of single pion events (CC1π) in the CCQE sample due to pion absorption in the nucleus"...and... " Also, there is a second band of events with $E_{TRUE}$ higher than $E_\nu^{QE}$ by about 300 MeV, which corresponds to CC1π events with an absorbed pion"[4]*.
[2] This information is not present in the tables I and II in [1] and can be extracted from the data release [5] for the "fully oscillated sample", including a ~2.4% contribution from the fluctuations of the decay length.

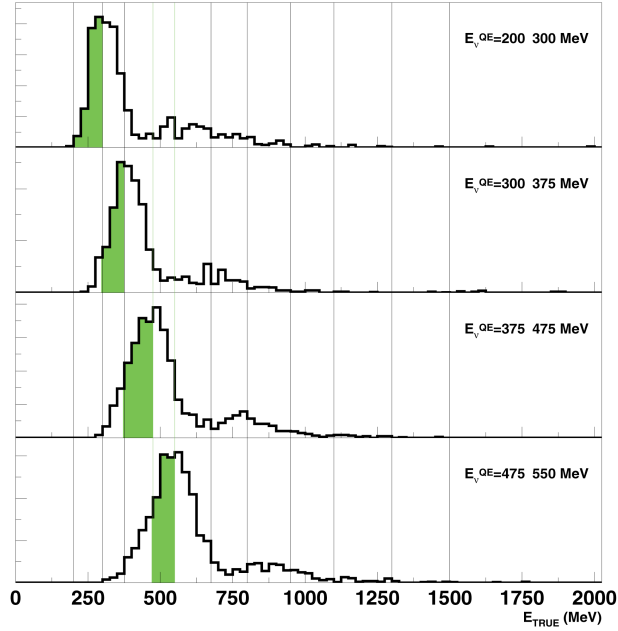

**Figure 3.** The spectrum of $E_{TRUE}$ separately for the first 4 bins in $E_\nu^{QE}$. The vertical lines correspond to the borders of the reconstructed energy bins adopted in MiniBooNE analysis, while the colored area corresponds to the events for which $E_{TRUE}$ overlaps to the corresponding bin of reconstructed energy $E_\nu^{QE}$.

| E bin | $E_\nu^{QE}$ | $E_{TRUE}$ | $L/E_\nu^{QE}$ | $L/E_{TRUE}$ | $E_\nu^{QE}/L$ | $E_{TRUE}/L$ |
|---|---|---|---|---|---|---|
| 200÷300 | 255±28 | 416±213 | 2.088±0.247 | 1.480±0.48 | 0,486±0,058 | 0.792±0.411 |
| 300÷375 | 341±20 | 465±187 | 1.545±0.107 | 1.246±0.33 | 0,650±0,046 | 0.888±0.363 |
| 375÷475 | 426±28 | 539±203 | 1.234±0.092 | 1.056±0.27 | 0,815±0,061 | 1.031±0.393 |
| 475÷550 | 513±20 | 607±178 | 1.021±0.051 | 0.920±0.22 | 0,982±0,051 | 1.163±0.347 |
| 550÷675 | 613±35 | 693±184 | 0.854±0.055 | 0.794±0.17 | 1,175±0,078 | 1.329±0.358 |
| 675÷800 | 737±35 | 793±173 | 0.707±0.040 | 0.685±0.15 | 1,419±0,082 | 1.529±0.341 |
| 800÷950 | 872±42 | 917±187 | 0.597±0.035 | 0.590±0.13 | 1,681±0,098 | 1.770±0.371 |
| 950÷1100 | 1021±43 | 1059±293 | 0.508±0.027 | 0.508±0.11 | 1,976±0,105 | 2.051±0.402 |
| 1100÷1300 | 1193±56 | 1203±216 | 0.434±0.025 | 0.445±0.10 | 2,311±0,134 | 2.332±0.429 |
| 1300÷1500 | 1388±57 | 1366±236 | 0.373±0.019 | 0.392±0.09 | 2,690±0,139 | 2.651±0.470 |
| 1500÷3000 | 1766±260 | 1666±386 | 0.298±0.037 | 0.331±0.10 | 3,411±0,509 | 3.229±0.753 |

**Table 1.** Modified version from Table I of Ref. [4], including the rms uncertainty on each column. Energies are measured in MeV, while the distance in meters.

According to the MiniBooNE paper [4], even after correcting for the substantial bias in E, the size of the smearing in the energy reconstruction makes it very difficult to represent the event excess as a function of L/E in an unambiguous way, or at least in an experiment independent way. Moreover establishing the most appropriate L/E value representing each experimental point is not straightforward, since it depends on the oscillation parameters (see f.i. table III and IV in Ref. [4]). The solution proposed by MiniBooNE in paper [4] to compare experimental data with theoretical oscillation probabilities requires:

- replacing the biased reconstructed energy with the expected average value of the

true neutrino energy (under the hypothesis of full oscillation of the muon neutrinos);
- comparing the observed $P_{osc}$ (defined as the observed event excess in each bin divided by the fully oscillated sample) with a smeared oscillation probability, to take into account the experimental effects in the energy reconstruction.

According to this approach, the smearing of the theoretical oscillation probability is determined by the experimental resolution but depends also on the oscillation parameters themselves, in particular the $\Delta m^2$ value, as shown in Figure 5.

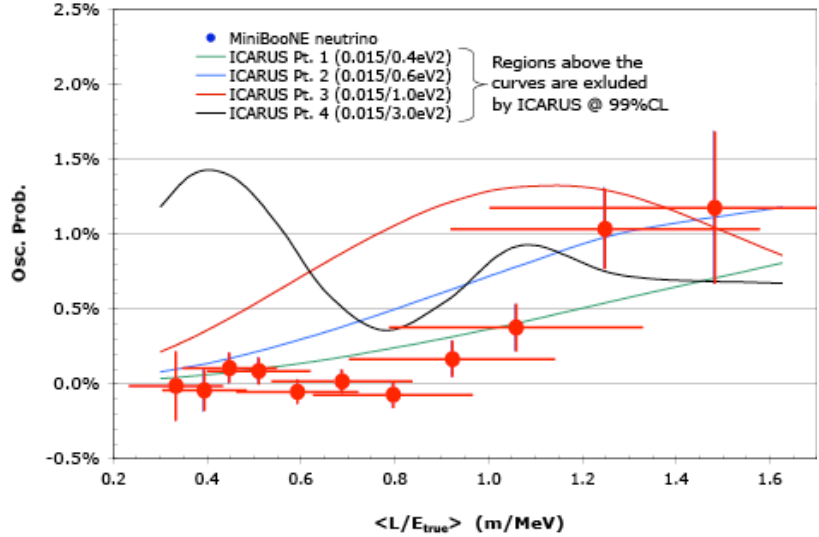

**Figure 4.** Modified version of Figure 7 in Ref. [1] showing the oscillation probability as a function of the average $L/E_{TRUE}$: the rms variation on $L/E_{TRUE}$ is associated to each point.

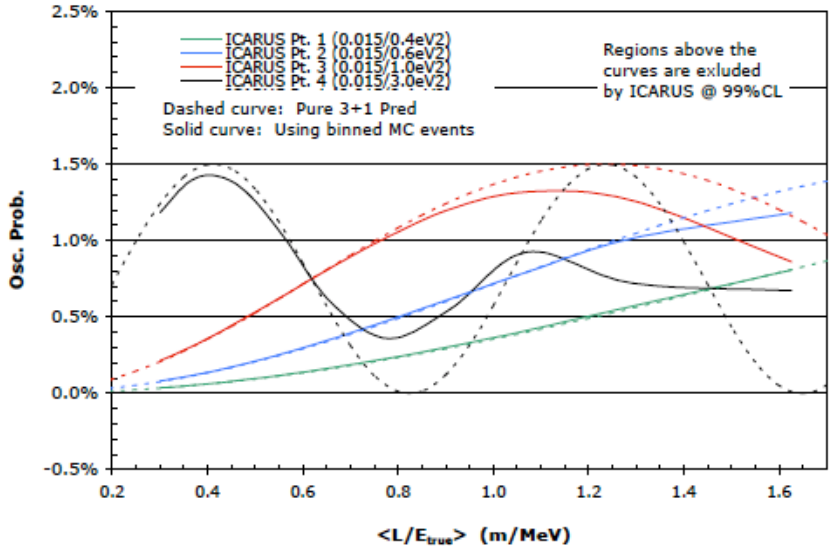

**Figure 5.** Reproduction of the Figure 6 from MiniBooNE paper [4] showing the predicted oscillation probability versus $L_{TRUE}/E_{TRUE}$ with (solid curves) and without (dashed curves) energy and flight path smearing for $\sin^2 2\theta = 0.015$ and $\Delta m^2 = 0.4, 0.6, 1.0$ and $3.0$ eV$^2$.

As a remark, the confusion between reconstructed energy and true energy in computing the oscillation probability is implicitly present also in an L/E plot of MiniBooNE paper [7] (Figure 6) where the LSND results are compared with the MiniBooNE data points. Here $P_{osc}$ is computed using the reconstructed energy $E_\nu^{QE}$ and not the neutrino energy. To remove the reconstructed energy bias, the MiniBooNE experimental points should be not negligibly moved to smaller $L/E_\nu$.

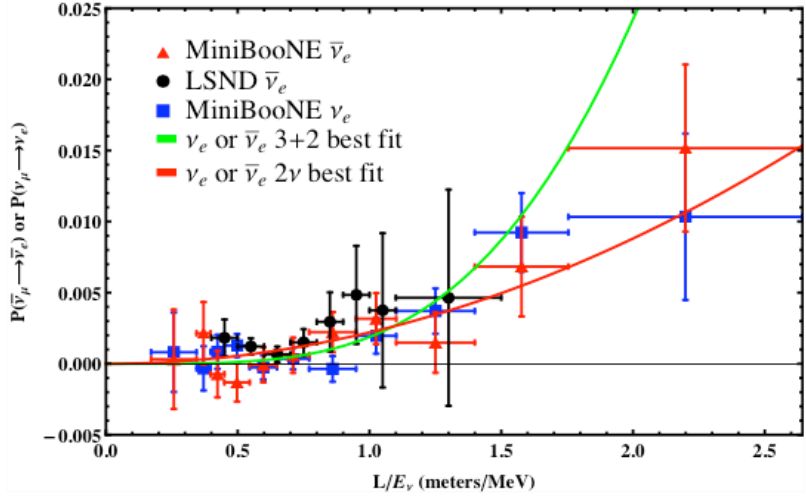

**Figure 6. The oscillation probability as a function of $L/E^{QE}$ for $\nu_\mu \rightarrow \nu_e$ and anti $\nu_\mu \rightarrow$ anti-$\nu_e$ (MiniBooNE) and anti-$\nu_\mu \rightarrow$ anti-$\nu_e$ (LSND) event candidates (the figure is reproduced from MiniBooNE paper [7]).**

The description of this figure in paper [7] is formally correct. Nevertheless, given the arguments of the last MiniBooNE paper [4] itself, it seems that this picture is conveying a misleading message:
- it legitimates a direct comparison of LSND results, for which the smearing is small and fairly Gaussian [4], with MiniBooNE for which the smearing is quite substantial and not Gaussian;
- it is comparing both LSND and MiniBooNE to the same theoretical predictions;
- it is using (for MiniBooNE) $L/E_\nu^{QE}$ which is shown in Ref. [4] to be strongly biased with respect to the actual $L/E_\nu$ value and poses strong difficulties in performing L/E comparisons.

According to the considerations of Ref. [4], for the sake of clarity and to avoid ambiguities, the MiniBooNE and LSND data should not be compared in the same figure.

As a second remark the MiniBooNE event excesses for both neutrino and antineutrino-modes in tables I and II in paper [4] have associated errors (which include both statistical and systematic uncertainties) which differ from the ones reported in Ref. [2] and somehow closer to the errors in 2012 arXiv paper [7], which was not published on a journal. The three sets of data are reported in Figure 7, illustrating the unexplained error variation from paper to paper and raising the relevant question of which errors should be used for the comparisons. Pending a clarification on this point, the present paper sticks to the last published results in Ref. [2].

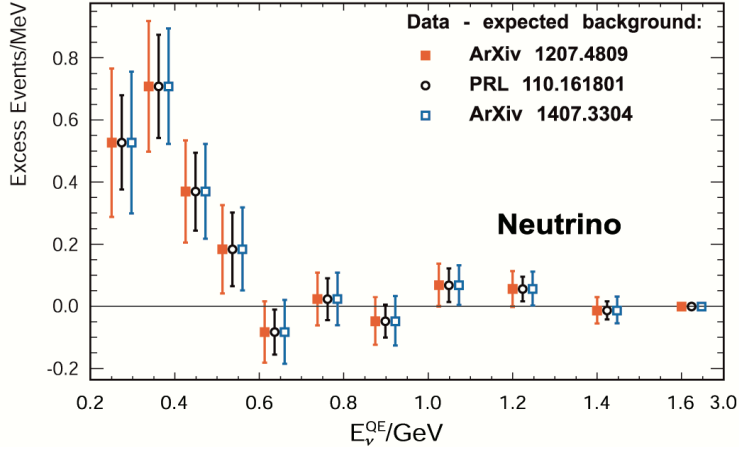

**Figure 7.** Comparison between the experimental errors quoted in the three quoted MiniBooNE papers [2,4,7]. The data sets in [2] and [4] are shifted horizontally to disentangle their vertical error bars.

## 3. Updating the ICARUS MiniBooNE comparison.

As a consequence of the large difference between $E_{TRUE}$ and $E_\nu^{QE}$ in MiniBooNE data quoted in Ref. [4], the interpretation of the MiniBooNE event excess in terms of oscillation probability as a function of L/E is troublesome, ambiguous and changing with the $\Delta m^2$ scale. The theoretical oscillation probability

$$P_{osc} = \sin^2 2\theta \cdot \sin^2(1.27 \Delta m^2 L / E)$$

is also affected in a complex way by the experimental smearing (as discussed in Ref. [4]) and cannot be simply superimposed on the experimental points in an L/E plot. To account for all these experimental effects, a viable alternative to the MiniBooNE procedure in Ref. [4] consists in binning both the data and the expected oscillation probability (using the same bins as the MiniBooNE analysis) and exploiting the MiniBooNE data release [5]. This allows reproducing the expected event excess in a two-neutrino oscillation scheme for any pair of oscillation parameters ($\Delta m^2$, $\sin^2(2\theta)$). As a consequence both the experimental "oscillation" signal $P_{osc}$ defined in each bin as:

$$(P_{osc}^{meas})_i = \frac{data_i - bkg_i}{fully\_osc_i}$$

and the associated theoretical predictions

$$(P_{osc}^{theor})_i = \frac{osc_i^{(\sin^2(2\theta), \Delta m^2)}}{fully\_osc_i}$$

can be drawn on the same plot. In these formulae $fully\_osc_i$ and $osc_i^{(\sin^2(2\theta), \Delta m^2)}$ are the expected contributions to the i[th] bin of $E_\nu^{QE}$, in case of full oscillation and of a two-neutrino oscillation with parameters ($\Delta m^2$, $\sin^2(2\theta)$) respectively. This approach has the advantage of correctly and coherently treating on the same footing both data and MC predictions. It fully accounts for the expected MiniBooNE detector response and for the binning chosen in the data representation. As already discussed, the LSND data cannot be accommodated in the same plot.

The result of such a treatment is shown in Figure 8. Within this new calculation framework the tension between the negative result of ICARUS experiment and the neutrino lowest energy points of MiniBooNE (200 < $E_\nu^{QE}$ < 475 MeV) is still observed for the 90% CL ICARUS limit, but is no more present for the 99% CL limit on $\sin^2(2\theta_{new})$. This new analysis is therefore confirming the suggestion, as already presented in Ref. [1], for an instrumental or otherwise unexplained nature of the low energy signal reported by MiniBooNE Collaboration in Ref. [2].

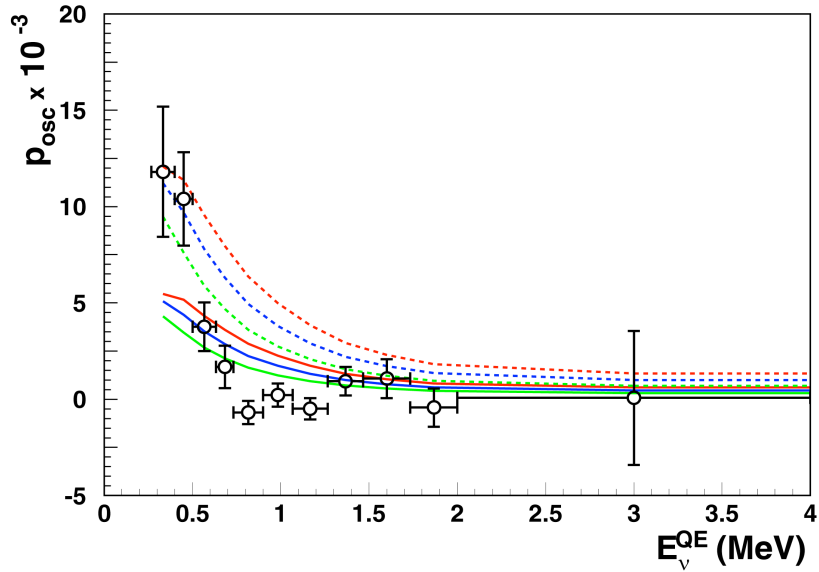

**Figure 8.** Updated version of figure 1: data points refer to the MiniBooNE experimental $P_{osc}$ revised following the adopted prescription (see text). The curves are the fit of the MC predicted $P_{osc}$ binned as the experimental data. The three continuous (dashed) curves correspond to $\sin^2(2\theta) = 0.0068$ ($\sin^2(2\theta) = 0.015$) i.e. the 90% CL (99% CL) ICARUS limit on the oscillation amplitude [1] and $\Delta m^2 = 0.5\ eV^2$ (green), 0.6 $eV^2$ (blue) and 0.7 $eV^2$ (red) respectively.

## References

[1] M. Antonello et al. [ICARUS Coll.], *Search for anomalies in the $\nu_e$ appearance from a $\nu_\mu$ beam*, Eur. Phys. J C73, 2599 (2013).

[2] A. A. Aguilar-Arevalo et al. [MiniBooNE Coll.], *Improved Search for anti $\nu_\mu \to$ anti $\nu_e$ Oscillation in the MiniBooNE Experiment*, Phys. Rev. Lett. 110, 161801 (2013).

[3] N. Agafonova et al. [OPERA Coll.], *Search for νμ→νe oscillations with the OPERA experiment in the CNGS beam*, JHEP 1307 (2013) 004.

[4] A. A. Aguilar-Arevalo et al., [MiniBooNE Coll.], *Using L/E Oscillation Probability Distributions*, arXiv:1407.3304v1 [hep-ex] (11 July 2014).

[5] http://www-boone.fnal.gov/for physicists/data release/nue nuebar 2012/.

[6] A. A. Aguilar-Arevalo et al., [MiniBooNE Coll.], *Search for Electron Neutrino Appearance at the $\Delta m^2 \sim 1\ eV^2$ scale*, Phys. Rev.Lett. 98, 231801 (2007)..

[7] A. A. Aguilar-Arevalo et al., [MiniBooNE Coll.], *A combined $\nu_\mu \to \nu_e$ & anti $\nu_\mu \to$ anti $\nu_e$ Oscillation Analysis of the MiniBooNE Excesses*, arXiv:1207.4809v2 (21 Aug 2012).